\documentclass[useAMS,usenatbib]{mn2e}
\usepackage{psfig}        

\usepackage[dvips]{graphicx}

\title[The search for the origin of the Local Bubble redivivus]{The
search for the origin of the Local Bubble redivivus}
\author[B. Fuchs, D. Breitschwerdt, M.A. de Avillez, C. Dettbarn, and C. Flynn]
{B. Fuchs$^{1}$\thanks{E-mail:fuchs@ari.uni-heidelberg.de}, D.
Breitschwerdt$^{2}$, M. A. de Avillez$^{2,3}$,
C. Dettbarn$^{1}$, and C. Flynn$^{4}$\\
$^{1}$Astronomisches Rechen-Institut, M\"onchhofstra{\ss}e 12-14, D-69120
Heidelberg, Germany\\
$^{2}$Institut f\"ur Astronomie der Universit\"at Wien,
T\"urkenschanzstra{\ss}e 17, A-1180 Wien, Austria\\
$^{3}$Department of Mathematics, University of \'Evora,
R. Rom\~ao Ramalho 59, 7000 \'Evora, Portugal \\
$^{4}$Tuorla Observatory, V\"ais\"al\"antie 20, FI-21500 Piikki\"o,
Finland}
\begin{document}

\date{Accepted  Received 2006}

\pagerange{\pageref{firstpage}--\pageref{lastpage}} \pubyear{2006}

\maketitle

\label{firstpage}

\begin{abstract}
We present a new unbiased search and analysis of all B stars in the
solar neighbourhood (within a volume of 400 pc diameter) using the
{\sf Arivel} data base to track down the remains of the OB
associations, which hosted the supernovae responsible for the Local
Bubble in the interstellar gas. We find after careful
dereddening and by comparison with theoretical isochrones, that besides the
Upper Scorpius the Upper Centaurus Lupus and Lower Centaurus Crux
subgroups are the youngest stellar associations in the solar
neighbourhood with ages of 20 to 30 Myr, in agreement with previous
work. In search for the ``smoking gun'' of the origin of the Local
Bubble, we have traced the paths of the associations back into the
past and found that they entered the present bubble region 10 to 15 Myr
ago. We argue that the Local Bubble began to form then and estimate that
14 to 20 supernovae have gone off since. It is shown that the implied energy
input is sufficient to excavate a bubble of the presently observed
size.
\end{abstract}


\section{Introduction}

The Local Bubble (LB), a low-density X--ray emitting cavity deficient
of H{\sc i}, is our Galactic habitat. Yet, until recently, its
origin remained mysterious. It was conjectured to be the result of
one or several supernova explosions (e.g.\ Cox \& Anderson 1982,
Innes \& Hartquist 1984, Smith \& Cox 2001), but firm evidence was
lacking, as no OB association was found within its boundaries,
extending about 200 pc in the Galactic plane, and 600 pc
perpendicular to it, but inclined by about $20^\circ$ with respect
to the axis of Galactic rotation, similar to Gould's Belt
(cf.\ Lallement et al.~2003). Further problems arose, since the X--ray and
EUV spectra measured in the Wisconsin Survey, by {\sf ROSAT PSPC, DXS,
XQC} and {\sf EUVE} were severely at odds with a thermal hot plasma in
collisional ionization equilibrium (CIE) as was pointed out by
Jelinsky et al.~(1995), Sanders et al.~(2001), McCammon et al.~(2002).
Most recently Hurwitz et al.~(2005) analyzed
{\sf CHIPS} data and found an extremely low emissivity of EUV iron lines.
The underabundance of soft X--ray lines can be naturally explained if
the plasma is in a state of delayed recombination (Breitschwerdt \&
Schmutzler 1994, Breitschwerdt 2001), but a high resolution
numerical hydrodynamical evolution model is needed to better
constrain non-equilibrium models. Spectral discrepancies between
models and observations can be alleviated if there is a substantial
contribution from very local sources, such as the Earth's exosphere
(Freyberg 1998) or charge exchange reactions between solar wind ions
(SWCE) and heliospheric gas (Lallement 2004). At present it
is unclear what fraction can be attributed to these very nearby
sources, although there is fairly robust evidence that even in the
extreme case of all of the X--ray emission being due to SWCE in a
certain direction, a substantial LB fraction remains, especially
perpendicular to the disc. For further details on LB properties we
refer to the review of Breitschwerdt (2001) and the conference
proceedings \emph{The Local Bubble and Beyond} (Breitschwerdt et al.~1998).

All these shortcomings have led several authors to
speculate that if the LB is not a classical superbubble, but rather an
appendix of the neighbouring Loop~I superbubble, which was expanding
into an interarm region  between the Sagittarius and the Perseus spiral arms
of the Galaxy (Bochkarev 1987, Frisch 1995). However, the
existence of a ``wall'' between the two bubbles, showing up in
absorption of soft X--rays in {\sf ROSAT PSPC} images (Egger \& Aschenbach
1995) renders this scenario not very plausible.

The search for the ``smoking gun'' of the origin of the LB proved
partially successful by discovering that moving groups of young
stars in the solar neighbourhood could provide an adequate number of
SN explosions while crossing the path of the LB. Bergh\"ofer \&
Breitschwerdt (2002, henceforth BB02) calculated the trajectory of
the Pleiades subgroup B1 backwards in time, and found that 19 SNe
could have exploded between 10--20 Myr ago in the region that is
occupied by the LB. The remaining stars of B1 are now part of the
Scorpius Centaurus OB association. It could be shown that this is in
good agreement with the size of the LB and the present soft X--ray
emissivity. A similar analysis was carried out by
Ma\'{i}z-Apell\'{a}niz (2001), who calculated backwards in time the
trajectories of Sco Cen subgroups and claimed that about 6 SNe that went
off in the Lower Centaurus Crux subgroup of the Sco OB2 association
7 to 9 Myr ago formed the LB.

While these analyses represent a major step towards the
understanding of the origin of the LB they are not free from bias,
in particular the assumption that \emph{certain stellar groups
should} be responsible for the sought SN explosions. The purpose of
this paper is to scrutinize \emph{all stars} that are within a
volume of about 400 pc in diameter centered around the Sun, and to
perform a selection according to spectral and kinematical
properties. The latter is based on three dimensional space velocities
of the stars. Thus our approach is complementary to studies like by
de Zeeuw et al. (1999) which are based on proper motions alone.
Sartori et al. (2003) do include radial velocities when analyzing the
subgroups of the Sco OB2 association, but work from a list of stars
preselected by de Zeeuw et al. (1999).
From their position in the HR diagram and the turnoff
point from the ZAMS, we can reliably determine the age of the stars
and estimate the number of SNe within a defined region, such as the
LB.

The paper is organized as follows. In the next section we describe
our search strategy for the remnants of the OB association
responsible for the origin of the LB. In section 3 we discuss the
consistency of our findings with the properties of the LB as
observed today, and present a high--resolution 3D hydrodynamical simulation
of the formation of the LB in the local interstellar gas. In the final
section we summarize our conclusions.

\section[]{Search for nearby OB associations}

Even though the Sco\,OB2 association has been claimed with good
reason to be responsible for the origin of the LB
(Ma\'{i}z-Apell\'{a}niz 2001, BB02), a fresh, unprejudiced search
for the OB association, that might have triggered the formation of
the LB, will improve and harmonize previous studies. As starting
point we used the {\sf Hipparcos} catalogue (ESA 1997) from which we
selected all stars bluer than $(B-V) < -0.05$ with parallaxes larger
than 5 milliarcseconds, because otherwise the distances would be too
inaccurate. Drimmel et al.~(2000) find by a comparison with the Tycho
catalogue that the {\sf Hipparcos} catalogue is about 97 percent
complete down to $V=7.5$ which corresponds at a distance of $1/(5mas)=200$ pc
to an absolute magnitude of $M_{\rm V}=1$. Allowing for an extinction of 
$A_{\rm V} \leq 0.2$ we have $M_{\rm V}<1.2$, and we have chosen the colour cut
in $(B-V)$ accordingly (cf. Figs. 4 and 5). We have obtained this way an 
unbiased complete sample of 762 B stars within a distance of 200 pc from the
Sun. One of the bluest stars is
the B0.5 star $\alpha$\,Cru. The stars of the sample show at this
stage already a well defined main sequence in the colour--magnitude
diagram (cf.~Fig.~4). There are a few stars lying several magnitudes
below the main sequence, which are probably subdwarfs. We have
omitted these, $M_{\rm V}>3.0$ mag, because we are interested in
young stars. All stars in our list appear in the {\sf Arivel} data
base, which combines {\sf Hipparcos} parallaxes and proper motions
with radial velocity data collected from the literature (Wielen et
al., in preparation). We find that for 610 stars radial velocities
are available. These are accurate enough to be used for a
kinematical analysis (cf.~Fig.~3). We have tested if the availability
of radial velocities is correlated with the colours of the stars and
find that the 152 out of 762 stars for which no radial velocity is known
lie above $(B-V) > -0.1$ . We discuss the implications of this effect on
the determination of the expected number of supernovae below.
For each star spatial $X$, $Y$,
and $Z$ coordinates and the corresponding $U$, $V$, and $W$ velocity
components have been calculated. The coordinates are centered on the
Sun with $X$ pointing towards $l=0^\circ$, $b=0^\circ$, $Y$ towards
$l= 90^\circ$, $b = 0^\circ$, and $Z$ towards $b = 90^\circ$,
respectively.
\begin{figure}
\centering
%
%
\caption[]{Positions of 610 stars drawn from
     the {\sf Hipparcos} catalogue. The selected stars have colours
     $B-V < -0.05$ and for each star its radial velocity is known. The
     $X$--axis points towards the Galactic centre, $Y$ into the direction of
    Galactic rotation, and $Z$ towards the Galactic north-pole, respectively.}
\label{fig:1}       
\end{figure}
Since OB associations disperse slowly on time scales of the order of several
$10^7$ years (Blaauw 1964), we have searched for kinematically coherent
structures in our sample.
For this purpose we have traced the positions of the stars backward
in time over 3$\cdot$10$^7$ years. Stellar orbits have been calculated using
the epicyclic equations of motion for the stars in the sample (Lindblad
1959, Wielen 1982)
\begin{eqnarray}
X(t) &=& X(0)-\frac{V(0)}{-2B}(1-\cos{(\kappa t)})+\frac{U(0)}{\kappa}
\sin{(\kappa t)}, \nonumber \\
U(t) &=& U(0)\cos{(\kappa t)}-\frac{\kappa}{-2B}V(0)\sin{(\kappa t)}, \nonumber
\\Y(t)& =& Y(0) + 2 A \left( X(0)-\frac{V(0)}{-2B} \right) t \\ & + &
 \frac{\Omega_0}{-B \kappa}V(0)\sin{(\kappa t)}  +
 \frac{2 \Omega_0}{\kappa^2}U(0) ( 1 -\cos{(\kappa t)}), \nonumber \\
V(t)& = &\frac{-2 B}{\kappa}U(0)\sin{(\kappa t)} + V(0)\cos{(\kappa t)},
\nonumber \\ Z(t) &=& \frac{W(0)}{\nu}\sin{(\nu t)} + Z(0)\cos{(\nu t)},
\nonumber \\ W(t)&=& W(0)\cos{(\nu t)} -Z(0)\nu \sin{(\nu t)}\,. \nonumber
\end{eqnarray}
In Eqns.~(1) $\kappa$ denotes the epicyclic frequency, $\kappa =
\sqrt{-4\Omega _0B}$. $A$ and $B$ are the Oort constants, and
$\Omega _0$ is the angular frequency of the rotation of the local standard
of rest around the Galactic centre, $ \Omega_0=V_{\rm LSR}/R_\odot$. $\nu$
denotes the vertical oscillation frequency which is related to the
local density $\rho_0$ by the Poisson equation  as $\nu= \sqrt{4\pi
G\rho_0}$, where $G$ is the constant of gravitation. For the angular velocity of
the local standard of rest we have adopted a value of 
$\Omega_0$ = 220 km/s/8kpc. The choice of the Oort constants was guided by
the consideration that they describe in eqns.~(1) the smooth Galactic 
gravitational potential. The latter is consistent with an essentially flat
shape of the local Galactic rotation curve, $A$ = $-B$ = $\Omega_0/2$
(Feast \& Whitelock 1997). This must not be confused with determinations of 
$A$ and $B$ using OB stars as, for instance, in the studies of Torra et
al.~(2000) or Elias et al.~(2006). These reflect peculiarities of the orbits of 
the OB stars in Gould's belt related to the velocities with which they were 
born, but not the characteristic smooth shape of the Galactic potential.
For the local density we adopt a value of $\rho_0$ = 
0.1 $\mathcal{M}_\odot$/pc$^3$ (Holmberg \& Flynn 2004). These parameter
values imply $\kappa$ = 0.039 km/s/pc = 4$\cdot$10$^{-8}$yr$^{-1}$
and $\nu$ = 0.074 km/s/pc = 7.5$\cdot$10$^{-8}$yr$^{-1}$. In Fig.~2
we show the positions of the stars today and 3$\cdot$10$^7$ years
ago. Apparently most stars came from directions $-90^\circ < l < 90^\circ$ 
and stayed close to the Galactic midplane. Most of the
610 stars do not belong to the OB association, which hosted the
SNe responsible for the origin of the LB, and have space
velocities different from the velocity of the association. Thus
they are dispersed away into a wide cloud. However, the overdense
regions in Fig.~2 indicate that there is a considerable number of
stars which stayed together. The larger size of the overdense
regions in the back projected sample compared to its size
today is obviously due to the observational errors. The
typical accuracy of {\sf Hipparcos} proper motions is about 1
mas/yr which corresponds at a distance of 100 pc to a velocity of
0.5 km/s, whereas the accuracy of the radial velocities is several
km/s. Taken together with an expansion velocity of the order of 10
km/s (Blaauw 1964), this implies a spreading of the overdensity,
which represents the kinematically homogenous group of stars, to a
size of roughly 500 $\times$ 500 pc in X and Y. As can be seen
from Fig.~2 there is an outer shroud of stars which lies at
greater distances from the core of the overdensity. These must be
stars with genuinely different space velocities from the
kinematically homogenous group of stars. We identify this
kinematically homogeneous group of stars as an OB association and
select 302 stars lying in the windows indicated as dashed lines in
Fig.~2. As expected these stars are more or less closely related
to the Sco\,OB2 association.
\begin{figure}
\centering
%
%
\caption[]{Positions of the originally selected stars today (pink) and
3$\cdot$10$^7$ years ago (blue). The Sun is at rest in the diagrams.
Stars lying in the windows indicated by dashed lines are identified as
putative members of the searched for OB association.}
\label{fig:2}       
\end{figure}

In Fig.~3 we show the present day velocity distribution of the 302
selected stars. Since the velocity dispersion of an OB association
is of the order of 10 km/s (Blaauw 1964) or even less (Kamaya 2004),
we make a second selection
indicated by windows drawn as dashed lines in Fig.~3. This leaves a
sample of 236 stars which we analyze in the following. {\sf
Hipparcos} numbers of these stars are listed in the appendix.
\begin{figure}
\centering
%
%
\caption[]{Present day velocity distribution of the 302 selected stars. A
second selection is made of the stars lying in the windows indicated by dashed
lines. }
\label{fig:3}       
\end{figure}

The final sample is shown as a colour--magnitude diagram in Fig.~4. For this
purpose we have cross--identified the sample stars in the {\sf Geneva
photometry} data base (Mermilliod, Hauck \& Mermilliod 2000) and replaced the
$(B-V)_T$ colours given in the {\sf Hipparcos} catalogue by $(B-V)_J$ colours,
because they can be then directly compared with theoretical isochrones available
in the literature. In the colour range, which we
consider here, $B-V$ given in the {\sf Tycho} system can not be transformed
directly to the Johnson system (ESA 1997). The absolute magnitudes have been
determined from the visual magnitudes given in the {\sf Hipparcos} catalogue
in the Johnson system.

We have compared our sample with the extensive membership list of
the Sco\,OB2 association compiled by de Zeeuw et al.~(1999)
who applied a combination of a modified convergent point method and the
so called spaghetti method (Hoogerwerf \& Aguilar 1999)
to {\sf Hipparcos} data. Of particular interest are the membership lists of the
subgroups Upper Scorpius (US), Upper Centaurus Lupus (UCL), and Lower
Centaurus Crux (LCC). With only very few exceptions all stars in the
membership lists, which
fulfill our colour selection criterion, appear also in our sample,
which gives confidence in our selection procedure. A few stars from
our final sample could be identified additionally in the membership
list of de Geus et al.~(1989) as members of the subgroups. The 79
stars common to both lists are colour coded in Fig.~4 and listed separately
in the appendix.
\begin{figure}
\centering
%
%
\caption[]{Colour--magnitude diagram of the final sample (236 stars). Members
of the Upper Centaurus Lupus subgroup of Sco\,OB2 are highlighted in orange,
Lower Centaurus Crux in yellow, and Upper Scorpius in grey, respectively.}
\label{fig:4}       
\end{figure}
\begin{figure}
\centering
%
%
\caption[]{Dereddened colour--magnitude diagram of the members of the Upper
Scorpius (grey), Upper Centaurus Lupus (orange) and Lower Centaurus Crux
(yellow) subgroups. The solid lines are theoretical isochrones colour coded
according to their ages.}
\label{fig:5}       
\end{figure}

\section[]{Results and discussion}
\subsection[]{The search for the ``smoking gun''}

The colour--magnitude diagram presented in Fig.~4 shows a clearly
discernible main sequence, which is particularly well delineated by
the members of the UCL group. The turn-off point
at the tip is defined by both the members of the UCL
and the LCC subgroups. Apparently these are together with the US subgroup
indeed the youngest OB associations in the solar neighbourhood
(de Geus et al. 1989, Sartori et al. 2003). In order to
determine their age we have compared the colour--magnitude diagram
with theoretical isochrones calculated by Schaller et al.~(1992) for
solar metallicities. Fortunately de Bruijne (1999) and Sartori et
al.~(2003) have determined individually for most members of the US, UCL, and
LCC subgroups, respectively, the extinction and colour excess by comparing the
observed $(V-I)_{\rm C}$ colours with the intrinsic colours of stars
of the same spectral type and luminosity class. Dereddened data of the 79 stars
are shown together with isochrones in Fig.~5. We conclude from Fig.~5 that
the ages of the UCL and LCC subgroups lie in the range of 20 to 30 Myr,
whereas we cannot date the age of the US subgroup on the basis of our data.
We note that this estimate of the ages of the subgroups is nearly twice
of that of de Geus et al.~(1989), who determined an age of 11 - 12 Myr of
the LCC subgroup and 14 - 15 Myr of the UCL subgroup, respectively. These
age estimates were revised by Sartori et al. (2003) to 16 - 20 Myr on
the basis of the Padova isochrones (Bertelli et al. 1994) instead of the
Maeder (1981a, b, c) isochrones, which were used by de Geus et al. (1989).
The Schaller et al. (1992) isochrones, which we used, are an upgrade of
Maeder's isochrones by the Geneva group. Moreover we note that Sartori
et al.~(2003) have adopted for the majority of their stars the spectral
types given in the {\sf Hipparcos} catalogue, which might not be as reliable as
the Geneva photometric data which we used. Given these uncertainties we
conclude that our age datings of the LCC and UCL subgroups are consistent
with Sartori et al.'s result. This agrees also well with
the age of Pleiades subgroup B1, which was suggested to be responsible for the
origin of the LB by BB02, but is significantly larger
than assumed by Ma\'{i}z-Apell\'{a}niz (2001), especially for the LCC
subgroup. Moreover, we have
examined with the help of the {\sf Simbad} data base each star of the
subgroups lying not on the main sequence and found that practically all these
stars are either binaries or peculiar in the sense that they are
variable, emission line stars etc.~(cf.~the notes to the tables), so
that their position off the main sequence in the colour--magnitude
diagram shown in Fig.~5 can be explained in our interpretation by
such effects.

In Fig.~6 we trace back the positions of the UCL
and LCC subgroup members over the last 30 Myr
using again the epicycle equations (1). However, we have not used
the individual space velocities of the stars, but adopted for each
star the mass--weighted mean velocity of the combined subgroups.
This avoids any unphysical spread of the spatial distribution of the
stars at earlier times due to the errors of the space velocity
components of the stars. The stellar masses have been determined
with the mass--to--magnitude relation
\begin{equation}
\mathcal{M}_\ast/\mathcal{M}_\odot = 3.857-1.453 M_{\rm V} +
0.183M_{\rm V}^2 + 0.069 M_{\rm v}^3 \,,
\end{equation}
which we have derived from a fit to Schaller et al.'s (1992)
isochrone data. To the mean velocity of the stars we have added the
solar motion $(U,V,W)_\odot = (10,5.3,7.2)$ km/s (Dehnen \& Binney
1998), so that the orbits are calculated in the reference frame of
the local standard of rest. Since interstellar gas has usually only
small peculiar motions, the local interstellar gas, and with it the
LB which is indicated in Fig.~6 by the contour line taken from Lallement et
al.~(2003), will basically corotate with the local standard of rest
around the Galactic centre. This means that the LB is at rest in the
reference frame of Fig.~6. As can been seen from Fig.~6 the path of
the association has aligned itself 15 Myr ago nearly parallel to
the tangential $Y$--direction, the direction of Galactic rotation.
Remnants of supernova explosions occurring during this period will
have experienced very little shear due to the differential rotation
of the Galaxy. The shear effect is described quantitatively by the
term linearly proportional to time in the epicycle equation for
$Y(t)$ (cf.~Eqs.~1). The coefficient $X(0)-\frac{V(0)}{-2B}$ is the
mean guiding centre radius of an orbit. If the spread of these radii
is small, as was the case in the last 15 Myr, the shear effect of
the Galactic differential rotation is minimized. In our view this
might well explain why all supernovae occurring during that time have
combined together to form the LB, while supernova remnants formed
at earlier times have drifted away (cf.~ Figs.~6 and 7).
\begin{figure}
\centering
%
%
\caption[]{Path of the Upper Centaurus Lupus and Lower Centaurus
Crux associations over the last 30 Myr projected onto the Galactic
plane. The look--back time is colour coded. The orbits are
calculated backwards in the reference frame of the local standard of
rest assuming for each star the same mass--weighted mean velocity of
the stars. The position of the Local Bubble is indicated by the
dash--dotted contour line and is at rest in this reference frame.}
\label{fig:6}       
\end{figure}

Next we illustrate in Fig.~7 the position of the Upper
Centaurus Lupus and Lower Centaurus Crux associations relative to
the LB today and at earlier times and reproduce the present day LB
contours in meridional sections through the bubble. From Fig.~6 we
estimate the Galactic longitude in which direction we expect the
associations to move. Choosing then the appropriate meridional
section through the bubble from the paper by Lallement et
al.~(2003), we can determine immediately the positions of the stars
in that longitude range relative to the LB. As can be seen from the
upper panels of Fig.~7 the associations are today just about to exit
the bubble. Five and 10 Myr ago they were inside. The bottom right panel
of Fig.~7 indicates that they entered 15 Myr ago the
region occupied by the LB today. In this scenario the LB was
starting to form about 15 Myr ago, which is consistent with the
estimates of the age of the LB by Ma\'{i}z-Apell\'{a}niz (2001) and
BB02. In this context it should be kept in mind that
although the contours determined by Lallement et al.~(2003) are the presently
best available, they are derived from Na{\sc i} absorption line
measurements, which allow to trace the  H{\sc i} distribution under
certain conditions, such as low temperatures ($< 10^4$ K), line
saturation for high column densities etc.~(see Sfeir et al.~1999 for a
discussion). In particular, the extension of the LB is \emph{unknown}
in directions where there are no background stars.
\begin{figure}
\centering
%
%
\caption[]{Meridional sections of the contours delineating the outer
boundary of the Local Bubble together with the positions of the stars in the
Upper Centaurus Lupus and Lower Centaurus Crux associations. The
horizontal axis in the upper left panel points into the direction
$l=300^\circ$, in the upper right towards $l=315^\circ$ and so on.
The vertical direction is always perpendicular to the Galactic midplane.
The ages of the associations are colour coded as in Fig.~6.}
\label{fig:7}       
\end{figure}
Regardless of the uncertainties of the outer boundary of the
LB one might wonder, moreover, how realistic a scenario is, in which the SNe
explode rather close to the edge of the present day bubble. As our
high resolution simulation discussed in the next section shows, the
location of the star cluster with respect to the centre of the
bubble is not crucial. The bubble expands always fastest in the
direction of the lowest ambient density and pressure. Since in the
direction of the Galactic centre the Loop~I superbubble was formed
almost at the same time as the LB by SNe exploding within the Sco--Cen
association, the pressure in this direction is very high. Hence the
LB was forced to expand rather towards the anticentre direction and
perpendicular to the plane, in agreement with the observations.

\subsection[]{Unravelling the supernovae of the Local Bubble}
Next we shall derive the number and masses of stars that
have exploded during the journey of the OB association through the
LB. Following BB02, the number of supernovae, which created the LB
can be estimated with the aid of the initial mass function (IMF). We
have fitted an IMF of the form (Massey et al.~1995)
\begin{equation}
\frac{d N }{d \mathcal{M}}=
\frac{d N }{d \mathcal{M}}\Big|_0
\mathcal{M}^{\Gamma -1}
\end{equation}
with an index $\Gamma = -1.1\pm 0.1$ to the data. Masses are given
in units of solar masses. The lower end of the main sequence at
$M_{\rm V} = 1$ mag corresponds to A0 stars with masses of
$\mathcal{M}_{\rm l} = 2.6\,\mathcal{M}_\odot$ and the upper tip at
$M_{\rm V} = -3.7$ mag to B0 stars with masses of $\mathcal{M}_{\rm
u} = 8.2\,\mathcal{M}_\odot$ (Schaller et al.~1992), respectively.
The total number of stars in the UCL and LCC associations, respectively,
allow the determination of the normalization constants,
\begin{equation}
N=\int_{2.6}^{8.2} \frac{d N }{d \mathcal{M}}\Big|_0
\mathcal{M}^{-2.1} d \mathcal{M}= \,
0.228\,\frac{d N }{d \mathcal{M}}\Big|_0 \,,
\end{equation}
implying $\frac{d N }{d \mathcal{M}}\big|_0$ = 184 for the 42 UCL
and $\frac{d N }{d \mathcal{M}}\big|_0$ = 118 for
the 27 LCC stars, respectively.
As we have shown in the previous section, OB stars entered the LB
region $10 - 15$ Myr ago, setting the clock for its origin to $t=0$.
From a further fit to Schaller et al.'s isochrone data we estimate
that the main sequence life time of such bright stars scales with
mass as
\begin{equation}
\tau=\tau_0\, {\mathcal{M}}^{-\alpha} \quad (2\,{\mathcal{M}}_\odot \leq
{\mathcal{M}}\leq 67\,{\mathcal{M}}_\odot )
\label{fuchs_formula}
\end{equation}
with $\tau _0 = 1.6 \cdot 10^{8 }$yr and $\alpha = 0.932$.
This means that the masses of the most massive stars
$\mathcal{M}_{\Delta \tau}$ in the associations at a lookback time of
$\Delta \tau $ years ago are given by
\begin{equation}
\mathcal{M}_{\Delta \tau}=\left( \mathcal{M}_{\rm u}^{-\alpha}-\frac{\Delta
\tau}{\tau_0} \right)^{-\frac{1}{\alpha}}\,,
\end{equation}
implying $\mathcal{M}_{10}=15.4\,\mathcal{M}_\odot $ if $\Delta
\tau= 10$ Myr or $\mathcal{M}_{15}=26.6\,\mathcal{M}_\odot $
if $\Delta \tau= 15$ Myr depending on the entry time of the associations into
the volume occupied by the LB today. The expected number of supernovae, i.e.
the number of 'missing' stars, is then calculated by
\begin{equation}
N_{\rm SN}=\int_{8.2}^{\mathcal{M}_{\Delta \tau}}
\frac{d N }{d \mathcal{M}}\Big|_0 \mathcal{M}^{-2.1} d \mathcal{M}\,.
\end{equation}
We thus obtain estimates of $N_{\rm SN} = 8 - 12$ from the Upper
Centaurus Lupus and $N_{\rm SN} = 6 - 8$ from the Lower Centaurus
Crux associations, respectively. The estimate of 14 to 20
supernovae, which created the LB,
is in good agreement with the values determined by Ma\'{i}z-Apell\'{a}niz 
(2001) and BB02. Extrapolating the IMF to masses beyond
${\mathcal{M}_{\Delta \tau}}$ we estimate that 12 to 5 SNe exploded before
the associations entered the present LB volume.

We have noted above that our original sample is complete in radial velocities
for stars with colours $(B-V) < -0.1$ which corresponds to $M_{\rm V}=0.7$,
if an extinction of $A_{\rm V}=0.1$ is assumed. According to Eq.~(2) such
stars  have a mass of 2.95 ${\mathcal{M}}_\odot$. If we remove 15 stars with
colours $(B-V) > -0.1$ from the final sample (79 stars) and modify Eq.~(4) for
the cut-off at 2.95  at the low mass end, we find $\frac{d N }
{d \mathcal{M}}\big|_{\rm 0, UCL+LCC}$ = 291 instead of 302. Thus the
incompleteness of the original sample has not introduced any significant bias
in our sample.

In order to assess the question, whether the estimated number of supernovae
would suffice to excavate the LB, we consider the energy input by
the supernova explosions into the interstellar gas. According to the
initial mass function (3) there are
\begin{equation}
d N = \frac{d N }{d \mathcal{M}}\Big|_0 \mathcal{M}^{\Gamma -1}
d\mathcal{M}
\label{starnumbers1}
\end{equation}
stars in the mass range $(\mathcal{M}, \mathcal{M}+d\mathcal{M})$ with
main sequence life times $(\tau ,\tau -d\tau )$. Thus
\begin{equation}
d N = \frac{d N }{d \mathcal{M}}\Big|_0 \mathcal{M}^{\Gamma -1}
\left(-\frac{d\mathcal{M}}{d\tau}\right)d\tau \,,
\label{starnumbers2}
\end{equation}
and the energy input rate is given by (cf.\ BB02)
\begin{eqnarray}
&& \dot{\mathcal{E}}_{\rm SN} = \frac{d}{dt}\mathcal{E}_{\rm SN} N_{\rm SN}
= \mathcal{E}_{\rm SN}\frac{dN_{\rm SN}}{dt} \\
&& =\mathcal{E}_{\rm SN}\frac{d N }{d \mathcal{M}}\Big|_0
\mathcal{M}^{\Gamma -1}(-1)\frac{d}{d\tau}
\left(\frac{\tau}{\tau_0}\right)^{-\frac{1}{\alpha}}\frac{d\tau}{dt}\,,
\nonumber
\label{en-input-rate1}
\end{eqnarray}
where $\mathcal{E}_{\rm SN}$ denotes the energy released by a single
supernova, $\mathcal{E}_{\rm SN} = 10^{51}$ ergs. According to the way we have
set up Eq.~(9) $\frac{d\tau }{dt }$ is equal to 1. Equation (10)
describes the trade--off of the increasing number of supernova progenitors and
their increasing main--sequence life times with decreasing mass. Inserting the
age--to--mass relation (5) into Eq.~(10) leads then to
\begin{equation}
\dot{\mathcal{E}}_{\rm SN} = \dot{\mathcal{E}}_{\rm
SN0}\,t_7^{-\frac{\Gamma +\alpha}{\alpha}}
\label{en-input-rate2}
\end{equation}
with $t_7$ defined as  $t_7 = t/10^7$yr. For the constant
$\dot{\mathcal{E}}_{\rm SN0}$ we find
\begin{eqnarray}
&& \dot{\mathcal{E}}_{\rm SN0} = \frac{\mathcal{E}_{\rm SN}}{\alpha \tau_0}
\frac{d N }{d \mathcal{M}}\Big|_0
\left(\frac{10^7 yr}{\tau_0}\right)^{- 0.1803} \\ \nonumber
&& = 3.5\cdot 10^{35}\frac{d N }{d \mathcal{M}}\Big|_0 {\rm erg/s}\,.
\nonumber
\label{en-input-rate3}
\end{eqnarray}
Equation~(\ref{en-input-rate2}) shows a rather weak decline of
the supernova energy input rate into the LB as a result of partial
compensation between the increasing number of stars with decreasing mass
and a corresponding increase in main sequence life time. It is quite
remarkable -- although probably fortuitous -- that the distribution
of stellar masses during the star formation process is nearly
anti-correlated with the main sequence life time of stars.

BB02 have derived a bubble wind equation which describes the growth
of the size of the bubble with time. The radius of the bubble is
given by
\begin{equation}
R_{\rm b}(t) = R_{\rm b0}\,t^{\frac{2 \alpha-\Gamma}{5 \alpha}}
\label{simsol}
\end{equation}
with the constant
\begin{equation}
R_{\rm b0}= \Big[  \frac{475
\alpha}{(4\alpha-7\Gamma)(3\alpha-4\Gamma)}\Big]^\frac{1}{5} \times
\Big[  \frac{\alpha\dot{\mathcal{E}}_{\rm SN0} \,
\tau_0^{1+\Gamma/\alpha}}{2 \pi (2-\Gamma)\rho_0}
\Big]^\frac{1}{5}\,.
\label{simsol_const}
\end{equation}
We note in passing that Eqs.~(13) and (14)
are consistent with Eqs.~(14) - (16) of BB02, except for a
different value of $\alpha$ used here, and a normalization error in
BB02, where $L_0$ and $\rho_0$ should be replaced by $\tilde L_0 =
L_0/t_0^\delta$ with $\delta=-(1+\Gamma/\alpha)$ and $\tilde \rho_0
= \rho_0/R_0^\beta$. In Eq.~(13) a constant density $\rho_0$ of
the ambient interstellar gas is assumed for which we adopt a value
of $\rho_0 = 2 \cdot 10^{-24}$ g/cm$^3$. The index in Eq.~(13),
($2\alpha  - \Gamma )/5\alpha = 0.564$, lies between the index of
0.4 of the Sedov equation, describing supernova remnants, and the
index of 0.6 of the stellar wind/superbubble expansion law. For a LB
age of 10 to 15 Myr Eq.~(13) predicts a bubble radius of 78~pc
to 100~pc, respectively. This in good agreement with the observed
size of the LB in the Galactic disk, as determined by Lallement et
al.~(2003; cf. also Fig.~7). For the determination of the expected
LB size we have used the expected numbers of supernovae both from the
LCC and UCL subgroups. Ma\'{i}z-Apell\'{a}niz (2001)
has argued that the LB owes its existence only to the 6 SNe stemming
from the LCC subgroup, because stars from this subgroup came closest
to the Sun in the past. We find the same when tracing the orbits of the
stars backwards in time. However, the members of the UCL subgroup did
enter the region occupied by the LB today and SNe stemming from the
UCL subgroup have to be taken into account, in our view, in the energy
considerations as well. The energy input of 6 SNe would excavate a bubble
with radius of only 65 pc, which is more difficult to reconcile with the fact
that the walls of the LB have been blown out above and below the
Galactic plane so that the LB has become effectively a chimney.
In general, however, similarity
solutions as applied here can only give a rough estimate of the LB age and
size due to several severe restrictions. Firstly, the ambient medium has to
be assumed to be either homogeneously distributed or to follow a power law
distribution in density, and its pressure has to be small compared to the
bubble pressure. Secondly, turbulent mixing and mass loading, which occur in
real bubbles, are hard to incorporate without further assumptions (cf.\ Dyson et
al.~2002). Therefore the most realistic approach to model
\emph{existing bubbles} is to perform 3D high--resolution numerical
simulations of their formation. A first simulation of this kind was
carried out by Breitschwerdt \& Avillez (2006) which was based on the older and
less detailed LB formation scenario of BB02. In the next section we present an
upgrade of that simulation which is now based on the better understood
supernova rate and the calculated paths of their progenitors through the LB as
derived in this paper.

\subsection{High Resolution Simulations of the Local Bubble Evolution}

We have simulated the effects of the explosions of the stars formerly
belonging to the UCL and LCC subgroups as their trajectories have crossed
the LB volume towards their present positions. The crucial
physical boundary conditions we have to apply to our simulations are the
locations, the masses, and derived from this, the explosion times of
the supernovae responsible for the origin of the LB. The latter can
be inferred from Eqs.~(\ref{starnumbers1}) and
(\ref{en-input-rate1}) to be
\begin{equation}
\frac{d N}{d\tau} = \frac{1}{\alpha \tau_0}\frac{d N }{d
\mathcal{M}}\Big|_0 \,
\left(\frac{\tau}{\tau_0}\right)^{-\frac{\Gamma + \alpha}{\alpha}}
\,, \label{snrate}
\end{equation}
which can be integrated to
\begin{equation}
N(\tau) = \frac{1}{\Gamma} \frac{d N }{d \mathcal{M}}\Big|_0 \,
\left[ \mathcal{M}^{\Gamma}_{\Delta \tau} -
\left(\frac{\tau}{\tau_0}\right)^{-\frac{\Gamma}{\alpha}}\right] \,,
\label{snnumber}
\end{equation}
taking the mass of the most massive star as an upper boundary.
We then proceed, somewhat arbitrarily, to bin the number of exploded
stars between $\mathcal{M}_u$ and $\mathcal{M}_{\Delta \tau}$ modulo
integer solar masses, and derive their main sequence and hence
explosion times from Eq.~(\ref{fuchs_formula}). Next, the explosion
locations are fixed by assuming that the presently ``missing stars''
were following the centres of mass of their respective subgroups.

The 3D high--resolution simulations are based on a hydrodynamical
Godunov scheme (cf.\ Godunov \& Ryabenki 1964) supplemented by
adaptive mesh refinement (AMR) along the lines described by Avillez \&
Breitschwerdt (2004), Breitschwerdt \& Avillez (2006). This entails a detailed
treatment of the evolution of the interstellar gas in a volume of
the Galaxy with a square area of 1 kpc$^{2}$ and a vertical extent
of 10 kpc on either side of the Galactic midplane based on the 3D
SN--driven ISM model of Avillez (2000) and Avillez \&
Breitschwerdt (2004). In these calculations the ISM is disturbed
by background supernova explosions at the Galactic rate. Initial
conditions for the \emph{ambient medium} were chosen from a data
cube of a previous hydrodynamical run where the highest adaptive
mesh refinement resolution was 1.25 pc (Avillez \& Breitschwerdt 2004,
Breitschwerdt \&  Avillez 2006). As a specific boundary condition we have to 
include the simultaneous evolution of the Loop~I superbubble, which has
been observed to interact with the Local Bubble according to ROSAT
PSPC observations (Egger and Aschenbach 1995). We therefore
selected a site with enough mass to form all the high mass stars
which are expected to explode as supernovae. Using the same IMF
for Galactic OB associations we derived in total 81 stars with
masses ${\mathcal{M}}$ between 7 and 31 ${\mathcal{M}}_\odot$
which in our simulations compose the Sco Cen cluster; 39 massive
stars with $14 \leq {\mathcal{M}} \leq 31 \,{\mathcal{M}}_\odot $
have already gone off, generating the Loop~I cavity (see Egger
1998, see also Avillez \& Breitschwerdt 2005a). Presently the Sco
Cen cluster, which is located at $(375,400)$ pc in the top panel
of Figure~\ref{lb1}, hosts 42 stars to explode within the next 13
Myr. Periodic boundary conditions are applied along the four
vertical boundary faces of our computational volume, while outflow
boundary conditions are imposed at the top ($z=10$ kpc) and bottom
($z=-10$ kpc) boundaries. The simulation time of this run was 30
Myr.

\begin{figure}
\centering
\caption[]{ Temperature (top panel) and pressure (bottom panel)
distributions in the Galactic midplane 13.4 Myr after the
first explosion in UCL occurred. The pressure is given in units of cm$^{-3}\,K$,
i.e.~divided by Boltzmann's constant $k$. The dimensions and morphology of
the Local Bubble are similar to the present observations. Loop~I,
to the right of the LB, is bounded by an X--ray illuminated shell
(top panel). \label{lb1}}
\end{figure}

Figure~\ref{lb1} shows the temperature (top) and pressure (bottom)
distributions in the Galactic midplane 13.4 Myr after the explosion of the
first SN, only a few thousand years after the last UCL and LCC supernovae
with masses of 8.2 ${\mathcal{M}}_\odot$ have exploded. This can be seen as a
red spot at ($x, y$) = (200, 300) pc. The LB is located
in the region between $100 \leq x\leq 300$ pc and $250 \leq y\leq 550$ pc,
its centre being located at $(x,y) = (200,400)$ pc. The shock waves of the
last two SNe occurring within the LB are most noticeable in the $P/k$
distribution by the high pressure peak shown in the bottom panel.
To the right of the LB the shell of Loop~I can be seen, which due
to its high temperature will emit in soft X--rays (top panel), consistent with
{\sf ROSAT PSPC} observations.

Another striking feature in Fig.~\ref{lb1} (bottom panel) are the coherent
bubble structures within a highly disturbed background medium with a pressure
in the range $2 \leq \log(P/k) \leq 4 $  which are due to the locally enhanced
SN rates in the vicinity of the Sun and in the Loop~I region. The
successive explosions close to the Sun heat and pressurize the LB,
which at first looks smooth, but develops internal temperature and
density structures at later stages. About 13.4 Myr after the first explosion
the LB cavity, which is bounded by an outer shell will start to fragment due
to Rayleigh--Taylor instabilities, in agreement with a linear stability
analysis carried out by Breitschwerdt et al. (2000). It then fills a volume
roughly corresponding to the present day LB size.

A more detailed analysis of these results and their observational
consequences will be the subject of forthcoming papers.

\section{Conclusions and outlook}

In contrast to previous analyses of the origin of the local
bubble we have not merely selected presently known stellar
subgroups and traced their kinematics back in time. Instead we
have scrutinized \emph{ab initio} a large sample volume of stars
for stellar groups by analyzing their spatial and kinematical
properties. From such an unbiased search among nearby B stars we
confirm the rather robust result that besides the Upper Scorpius
(US) subgroup the Upper Centaurus Lupus (UCL), and Lower Centaurus
Crux (LCC) subgroups with ages of 20 - 30 Myr are the youngest
stellar associations in the solar neighbourhood. Our search
volume is presently limited to a diameter of 400 pc, because the
{\sf Hipparcos} parallaxes are not accurate enough at distances
larger than 200 pc. Hence the analysis of a larger volume has to
await the launch of {\sf GAIA}.

Our search strategy relied mainly on kinematical criteria, and we
found many other B stars with the same kinematics as the
subgroups.  We have followed the paths of the associations into
the past and find that they entered the region of the present LB
10 to 15 Myr ago. Deriving O{\sc vi} column densities from a
numerical simulation of the general ISM (Avillez  \& Breitschwerdt
2005b) as well as of LB and Loop~I evolution (Breitschwerdt \&
Avillez 2006) in a realistic background medium, excellent
agreement was found with O{\sc vi} absorption line data obtained
with {\sf FUSE} (Oegerle et al.~2005, Savage \& Lehner 2006).
According to numerical LB evolution simulations by Breitschwerdt
\& Avillez (2006), who used supernovae from the subgroup B1 of the
Pleiades to power the LB, the O{\sc vi} data can be fitted with a
LB age of $14.4\pm^{0.7}_{0.4}$ Myr. The age of 13.9 - 14.1 Myr
estimated from the present simulation is thus consistent with the
age estimated from the slightly different simulation by
Breitschwerdt \& Avillez (2006). We therefore conclude that the LB
must have been excavated during this time. We find that about 14
to 20 SNe originated from the associations LCC and UCL. The
implied energy input into the ambient interstellar gas explains
quantitatively the present size of the LB.

The LB serves as an ideal test laboratory for superbubble
models due to the wealth of observations against which they can be
tested. Apart from the important O{\sc vi} test, we will also
compare EUV and soft X--ray emission data with our models in order to
derive the excitation history of ions in the LB and a possible
deviation from collisional ionisation equilibrium.

\section*{Acknowledgments}

This research has made extensive use of the {\sf Simbad} data
base at CDS, Strasbourg, France. This work has been partially funded
by the Portuguese Science Foundation under the project
PESO/P/PRO/40149/2000 to MAdeA and DB. CF thanks the Academy of Finland for
funding a one--month stay in Germany during which part of this work was carried
out. We thank Verena Baumgartner for careful reading of the manuscript.

\appendix
\section[]{{\sf Hipparcos} numbers of selected stars}
\begin{table*}
\hspace*{-3.5cm}
\centering
 \begin{minipage}{140mm}

  \caption{Identified members of Upper Scorpius (ass=1), Upper Centaurus
  Lupus (ass=2), and Lower Centaurus Crux (ass=3). Positional and velocity
  errors are given by the $\epsilon_{i}$.}
  \begin{tabular}{@{}lrrrrrrrrrrrrrrr@{}}
  \hline
  HIP- & ass & $M_{V}$ & $(B-V)_0$ & X & Y & Z & U & V & W &
$\epsilon_{X}$ & $\epsilon_{Y}$ & $\epsilon_{Z}$ &
$\epsilon_{U}$ & $\epsilon_{V}$ & $\epsilon_{W}$ \\
   no. &  & [mag] & [mag] & [pc] & [pc] & [pc] & [km/s]& [km/s]& [km/s]&
 &  &  &  &  &  \\ \hline
 50847$^{13}$ & 3 &-0.63 & -0.132 &  42.7  & -123.3 & -18.6  & -11.06 & -16.11 &  -3.62 &  2.8 &  8.1 &  1.2 &  1.5 &  3.5 &  0.6 \\
 53701$^{15}$ & 3 & 0.82 & -0.098 &  37.8  & -104.6 &  -2.6  &  -8.03 & -19.46 &  -5.77 &  2.5 &  6.9 &  0.2 &  1.6 &  3.5 &  0.6 \\
 55425$^{15}$ & 3 &-1.07 & -0.157 &  33.4  &  -92.1 &  10.4  & -10.80 & -14.61 &  -5.61 &  1.8 &  4.8 &  0.6 &  1.5 &  3.5 &  0.6 \\
 57851$^{  }$ & 3 &-0.31 & -0.156 &  46.6  &  -92.5 &  -5.5  &  -4.95 & -25.11 &  -8.09 &  2.7 &  5.4 &  0.3 &  1.2 &  1.6 &  0.5 \\
 58326$^{  }$ & 3 &-0.73 & -0.157 &  82.6  & -163.8 &  -0.7  &  -8.65 & -26.16 &  -8.81 &  7.9 & 15.6 &  0.1 &  2.4 &  3.4 &  1.0 \\
 58720$^{  }$ & 3 & 1.01 & -0.080 &  45.0  &  -82.5 & -11.1  & -11.33 & -13.92 &  -7.20 &  2.3 &  4.2 &  0.6 &  1.9 &  3.2 &  0.6 \\
 59173$^{  }$ & 3 &-0.94 & -0.192 &  49.4  & -101.5 &  23.2  &  -7.84 & -23.69 &  -4.98 &  4.1 &  8.4 &  1.9 &  1.5 &  1.8 &  0.9 \\
 59449$^{  }$ & 3 &-1.17 & -0.171 &  46.6  &  -92.2 &  18.3  &  -9.60 & -23.79 &  -9.51 &  3.6 &  7.1 &  1.4 &  2.1 &  3.4 &  1.2 \\
 59747$^{  }$ & 3 &-2.45 & -0.237 &  52.7  &  -98.1 &   7.4  &  -5.49 & -28.71 &  -6.86 &  3.5 &  6.6 &  0.5 &  1.4 &  1.7 &  0.6 \\
 60009$^{  }$ & 3 &-1.20 & -0.187 &  54.2  &  -96.5 &  -2.6  &  -6.81 & -21.74 &  -8.00 &  3.3 &  5.9 &  0.2 &  1.0 &  0.8 &  0.6 \\
 60710$^{  }$ & 3 &-0.68 & -0.162 &  58.2  & -105.2 &  23.9  & -12.13 & -13.98 &  -6.34 &  4.9 &  8.9 &  2.0 &  2.2 &  3.3 &  1.0 \\
 60823$^{  }$ & 3 &-1.76 & -0.198 &  64.5  & -115.9 &  29.3  & -12.72 & -18.47 &  -7.94 &  5.9 & 10.6 &  2.7 &  2.3 &  3.3 &  1.3 \\
 61585$^{  }$ & 3 &-2.17 & -0.212 &  48.9  &  -79.3 & -10.3  &  -8.39 & -19.51 &  -7.89 &  2.2 &  3.6 &  0.5 &  2.0 &  3.2 &  0.5 \\
 62058$^{  }$ & 3 & 0.47 & -0.079 &  66.4  & -107.2 &  14.8  &  -9.90 & -19.18 &  -6.21 &  5.4 &  8.7 &  1.2 &  1.5 &  1.7 &  0.7 \\
 62327$^{  }$ & 3 &-0.86 & -0.183 &  64.3  & -102.0 &  13.5  &  -6.65 & -24.85 &  -6.89 &  4.7 &  7.4 &  1.0 &  1.5 &  1.7 &  0.7 \\
 62434$^{16}$ & 3 &-3.92 & -0.240 &  57.9  &  -91.1 &   6.0  & -12.21 & -26.68 &  -6.05 &  3.8 &  6.0 &  0.4 &  1.5 &  1.1 &  0.6 \\
 63003$^{  }$ & 3 &-1.29 & -0.180 &  63.3  &  -96.2 &  11.5  &  -6.03 & -21.39 &  -5.81 &  4.2 &  6.5 &  0.8 &  1.0 &  0.9 &  0.6 \\
 63005$^{  }$ & 3 &-0.29 & -0.148 &  60.6  &  -92.0 &  11.0  &  -6.83 & -20.64 &  -4.21 &  4.1 &  6.2 &  0.7 &  2.2 &  3.2 &  0.6 \\
 63007$^{  }$ & 3 &-0.63 & -0.166 &  60.5  &  -92.0 &   7.2  &  -7.81 & -20.05 &  -6.70 &  4.0 &  6.1 &  0.5 &  2.2 &  3.2 &  0.6 \\
 63945$^{  }$ & 3 &-0.83 & -0.147 &  71.3  & -100.1 &  31.4  & -10.48 & -16.96 &  -6.03 &  6.5 &  9.1 &  2.9 &  2.4 &  3.1 &  1.2 \\
 64004$^{13}$ & 3 &-1.35 & -0.227 &  71.5  & -100.2 &  28.2  &  -4.21 & -21.89 &  -3.11 &  7.3 & 10.2 &  2.9 &  2.5 &  3.1 &  1.1 \\
 64053$^{  }$ & 3 & 0.64 & -0.093 &  56.8  &  -80.0 &  16.1  &  -1.27 & -29.22 &  -4.81 &  3.7 &  5.2 &  1.0 &  2.3 &  3.1 &  0.9 \\
 64425$^{18}$ & 3 &-0.56 & -0.081 &  61.6  &  -86.3 &   5.3  &  -5.04 & -18.79 &  -6.99 &  9.9 & 13.9 &  0.9 &  2.9 &  3.4 &  1.4 \\
 65112$^{  }$ & 3 & 0.05 & -0.132 &  72.6  &  -95.0 &  20.8  &  -8.05 & -20.15 &  -5.56 &  5.9 &  7.7 &  1.7 &  2.5 &  3.1 &  0.9 \\
 65271$^{  }$ & 3 &-0.78 & -0.173 &  64.9  &  -87.1 &   3.1  & -11.63 & -16.36 &  -5.38 &  4.1 &  5.5 &  0.2 &  2.4 &  3.1 &  0.4 \\
 66454$^{  }$ & 3 & 0.52 & -0.112 &  75.7  &  -86.6 &  32.4  &  -7.46 & -17.85 &  -3.78 &  6.9 &  7.9 &  2.9 &  2.3 &  2.6 &  1.1 \\
 67036$^{03}$ & 3 & 1.18 &  0.098 &  73.0  &  -83.0 &  21.5  &  -6.08 & -19.46 &  -5.23 &  6.2 &  7.0 &  1.8 &  1.5 &  1.7 &  0.8 \\
 67464$^{13}$ & 2 &-2.40 & -0.229 &  95.8  &  -97.8 &  49.5  &  -7.33 & -22.19 &  -6.12 & 10.7 & 11.0 &  5.6 &  1.5 &  2.0 &  1.1 \\
 67472$^{05}$ & 2 &-2.58 & -0.180 & 106.5  & -109.4 &  52.9  &  -6.86 & -23.38 &  -6.42 & 12.2 & 12.5 &  6.1 &  2.4 &  2.7 &  1.4 \\
 67669$^{  }$ & 2 &-0.54 & -0.149 &  59.1  &  -54.6 &  43.1  &  -4.63 & -20.69 &  -3.96 &  4.7 &  4.4 &  3.5 &  1.3 &  1.6 &  1.1 \\
 67973$^{  }$ & 2 & 0.51 & -0.090 &  70.2  &  -75.9 &  17.3  &  -0.22 & -24.56 &  -3.88 &  4.8 &  5.2 &  1.2 &  3.4 &  3.7 &  1.0 \\
 68245$^{  }$ & 2 &-1.94 & -0.219 &  96.9  &  -93.7 &  46.6  &  -7.62 & -19.12 &  -6.37 &  9.3 &  9.0 &  4.5 &  1.5 &  1.8 &  1.0 \\
 68282$^{  }$ & 2 &-1.68 & -0.209 &  87.3  &  -86.4 &  36.3  &  -9.09 & -18.99 &  -6.43 &  7.8 &  7.7 &  3.2 &  1.6 &  1.8 &  0.9 \\
 68862$^{  }$ & 2 &-1.33 & -0.199 &  95.4  &  -86.7 &  45.8  &  -4.18 & -22.32 &  -4.88 &  9.8 &  8.9 &  4.7 &  2.3 &  2.5 &  1.4 \\
 69618$^{06}$ & 2 &-1.07 & -0.138 & 103.5  & -106.7 &  10.3  &  -9.63 & -19.90 &  -8.16 &  9.1 &  9.4 &  0.9 &  5.3 &  5.5 &  1.0 \\
 70300$^{  }$ & 2 &-1.16 & -0.199 &  94.5  &  -75.0 &  44.0  &  -7.24 & -18.06 &  -5.30 &  9.2 &  7.3 &  4.3 &  1.5 &  1.8 &  0.8 \\
 70455$^{  }$ & 2 & 0.92 & -0.089 & 115.1  &  -91.9 &  50.0  &  -1.80 & -24.44 &  -3.46 & 15.4 & 12.3 &  6.7 &  4.7 &  4.2 &  2.2 \\
 70626$^{  }$ & 2 & 0.62 & -0.087 & 104.6  &  -81.3 &  46.7  &  -3.76 & -19.72 &  -6.36 & 11.6 &  9.0 &  5.2 &  2.1 &  2.4 &  1.4 \\
 71352$^{07}$ & 2 &-2.73 & -0.270 &  72.2  &  -54.8 &  27.1  & -10.77 & -17.42 &  -6.55 &  5.7 &  4.3 &  2.1 &  1.5 &  1.7 &  0.8 \\
 71453$^{  }$ & 2 & 0.19 & -0.117 &  98.5  &  -72.0 &  40.5  &  -7.53 & -16.98 &  -5.52 &  9.8 &  7.1 &  4.0 &  3.0 &  2.6 &  1.4 \\
 71536$^{03}$ & 2 &-0.84 & -0.150 &  72.0  &  -60.1 &  16.3  &  -3.79 & -18.67 &  -5.44 &  4.9 &  4.1 &  1.1 &  5.6 &  4.8 &  1.4 \\
 71724$^{15}$ & 2 & 1.16 & -0.086 &  94.9  &  -68.5 &  36.9  &  -5.83 & -15.03 &  -6.25 &  9.7 &  7.0 &  3.8 &  1.1 &  1.6 &  1.0 \\
 71727$^{15}$ & 2 & 0.57 & -0.124 & 123.7  &  -96.5 &  36.2  &  -9.74 & -20.50 &  -8.01 & 16.9 & 13.2 &  5.0 &  5.9 &  5.1 &  2.2 \\
 71860$^{09}$ & 2 &-3.87 & -0.222 & 129.1  & -102.3 &  33.3  &  -8.77 & -22.92 &  -9.13 & 16.5 & 13.1 &  4.3 &  1.7 &  2.5 &  1.6 \\
 71865$^{  }$ & 2 &-0.88 & -0.187 &  72.7  &  -49.2 &  32.1  &  -6.09 & -17.31 &  -5.17 &  4.9 &  3.3 &  2.2 &  1.2 &  1.3 &  0.7 \\
 72683$^{10}$ & 2 &-1.21 & -0.167 &  99.2  &  -69.7 &  30.5  &  -4.44 & -21.75 &  -5.35 &  9.7 &  6.8 &  3.0 &  1.1 &  1.7 &  0.8 \\
 72800$^{  }$ & 2 &-0.34 & -0.167 &  94.3  &  -59.1 &  38.6  &  -2.09 & -16.26 &  -1.16 &  8.4 &  5.3 &  3.5 &  1.6 &  1.5 &  0.7 \\
 73334$^{  }$ & 2 &-2.99 & -0.216 & 133.9  &  -87.4 &  42.1  &  -2.94 & -22.31 &  -5.53 & 16.1 & 10.5 &  5.1 &  1.3 &  2.1 &  1.0 \\
 73807$^{15}$ & 2 &-2.01 & -0.186 & 123.5  &  -85.4 &  26.3  &  -8.33 & -21.29 &  -3.97 & 18.1 & 12.5 &  3.8 &  3.5 &  3.4 &  1.2 \\
 74066$^{15}$ & 2 & 0.24 & -0.150 & 104.9  &  -62.6 &  33.2  &  -8.53 & -23.00 &  -5.73 & 11.3 &  6.7 &  3.6 &  6.2 &  4.3 &  2.1 \\
 74100$^{  }$ & 2 & 0.18 & -0.118 & 112.5  &  -70.2 &  31.1  &  -7.15 & -17.28 &  -4.37 & 12.6 &  7.8 &  3.5 &  3.4 &  2.7 &  1.3 \\
 74479$^{15}$ & 2 & 0.83 & -0.091 &  93.5  &  -48.6 &  35.3  &  -8.96 & -14.13 &  -4.21 &  8.6 &  4.5 &  3.2 &  3.2 &  2.2 &  1.3 \\
 74950$^{11}$ & 2 &-0.40 & -0.103 & 133.7  &  -74.6 &  38.0  &   0.26 & -24.24 &  -2.81 & 15.2 &  8.5 &  4.3 &  6.3 &  4.1 &  2.0 \\
 75141$^{13}$ & 2 &-2.79 & -0.238 & 133.3  &  -72.9 &  37.4  & -11.02 & -18.73 &  -7.28 & 17.9 &  9.8 &  5.0 &  2.7 &  3.0 &  1.2 \\
 75151$^{15}$ & 2 & 1.11 & -0.113 & 105.6  &  -54.4 &  33.6  &  -8.85 & -13.54 &  -5.13 & 14.6 &  7.5 &  4.7 &  5.5 &  3.4 &  1.9 \\
 75264$^{  }$ & 2 &-2.61 & -0.192 & 130.6  &  -77.8 &  27.7  &  -3.20 & -21.98 &  -2.56 & 12.3 &  7.3 &  2.6 &  3.3 &  2.6 &  0.9 \\

\hline
\end{tabular}\\
\end{minipage}
\end{table*}

\addtocounter{table}{-1}
\begin{table*}
\hspace*{-3.5cm}
 \centering
 \begin{minipage}{140mm}
  \caption{continued}
  \begin{tabular}{@{}lrrrrrrrrrrrrrrr@{}}
  \hline
  HIP- & ass & $M_{V}$ & $(B-V)_0$ & X & Y & Z & U & V & W &
$\epsilon_{X}$ & $\epsilon_{Y}$ & $\epsilon_{Z}$ &
$\epsilon_{U}$ & $\epsilon_{V}$ & $\epsilon_{W}$ \\
   no. &  & [mag] & [mag] & [pc] & [pc] & [pc] & [km/s]& [km/s]& [km/s]&
 &  &  &  &  &  \\ \hline
 75304$^{  }$ & 2 &-1.83 & -0.158 & 159.8  &  -78.5 &  53.6  &  -8.17 & -22.61 &  -5.98 & 22.9 & 11.2 &  7.7 &  2.7 &  3.5 &  1.3 \\
 75647$^{  }$ & 2 &-0.08 & -0.148 & 111.3  &  -52.8 &  36.1  &  -7.74 & -17.37 &  -4.39 & 12.0 &  5.7 &  3.9 &  8.7 &  4.6 &  2.9 \\
 76297$^{  }$ & 2 &-3.41 & -0.217 & 151.9  &  -76.7 &  35.8  &  -8.26 & -21.78 &  -8.70 & 32.8 & 16.5 &  7.7 &  8.9 &  6.4 &  2.9 \\
 76371$^{  }$ & 2 &-1.07 & -0.184 & 115.1  &  -63.8 &  20.3  &  -2.86 & -19.47 &  -2.27 & 11.2 &  6.2 &  2.0 &  1.7 &  1.8 &  0.6 \\
 76395$^{  }$ & 2 & 0.95 & -0.103 &  99.4  &  -39.3 &  35.1  &  -8.24 & -14.60 &  -4.54 &  9.3 &  3.7 &  3.3 &  3.3 &  2.0 &  1.3 \\
 76945$^{  }$ & 2 &-0.67 & -0.145 & 108.5  &  -42.8 &  33.6  &  -3.21 & -22.09 &  -3.65 & 11.1 &  4.4 &  3.4 &  1.7 &  2.2 &  1.0 \\
 77286$^{  }$ & 2 & 0.23 & -0.118 & 107.2  &  -40.8 &  32.0  &  -0.87 & -19.16 &  -2.89 & 11.6 &  4.4 &  3.5 &  6.7 &  3.2 &  2.1 \\
 77635$^{  }$ & 1 &-1.77 & -0.170 & 144.3  &  -35.7 &  59.2  &  -5.12 & -19.74 &  -7.55 & 21.0 &  5.2 &  8.6 &  4.3 &  3.0 &  2.0 \\
 77840$^{01}$ & 1 &-1.31 & -0.160 & 120.4  &  -28.1 &  49.0  & -10.45 & -15.61 &  -8.52 & 18.9 &  4.4 &  7.7 &  1.5 &  2.6 &  1.1 \\
 77900$^{  }$ & 1 &-0.01 & -0.096 & 148.2  &  -38.1 &  55.7  &  -3.92 & -21.17 &  -7.88 & 19.5 &  5.0 &  7.3 &  2.4 &  3.0 &  1.3 \\
 77909$^{  }$ & 1 & 0.04 & -0.097 & 126.7  &  -29.0 &  51.2  &  -9.40 & -15.78 &  -9.41 & 21.2 &  4.8 &  8.6 &  3.4 &  2.8 &  1.6 \\
 78207$^{02}$ & 1 &-1.03 & -0.089 & 137.7  &   -8.7 &  75.3  &  -6.74 & -15.01 &  -5.07 & 17.1 &  1.1 &  9.4 &  1.6 &  2.0 &  1.0 \\
 78246$^{13}$ & 1 &-0.60 & -0.140 & 136.6  &  -28.7 &  54.1  & -12.34 & -15.49 & -10.56 & 16.8 &  3.5 &  6.6 &  3.1 &  2.2 &  1.5 \\
 78265$^{13}$ & 1 &-2.85 & -0.249 & 128.9  &  -29.2 &  48.7  & -12.31 & -15.32 & -10.47 & 15.2 &  3.5 &  5.8 &  6.8 &  2.6 &  2.7 \\
 78384$^{  }$ & 2 &-2.48 & -0.226 & 138.4  &  -53.8 &  28.9  &  -5.80 & -21.12 &  -6.28 & 16.3 &  6.3 &  3.4 &  3.6 &  2.8 &  1.2 \\
 78655$^{  }$ & 2 &-1.14 & -0.141 & 148.4  &  -56.6 &  29.2  &  -9.36 & -22.89 &  -7.30 & 18.2 &  6.9 &  3.6 &  2.8 &  3.0 &  1.1 \\
 78756$^{15}$ & 2 & 0.83 & -0.056 & 153.9  &  -59.9 &  28.0  &  -6.36 & -22.90 &  -4.40 & 24.5 &  9.5 &  4.5 &  6.9 &  4.5 &  1.7 \\
 78877$^{04}$ & 1 &-0.02 & -0.096 & 137.5  &  -23.4 &  53.1  &  -6.35 & -19.40 & -10.64 & 17.6 &  3.0 &  6.8 &  3.4 &  2.7 &  1.7 \\
 79044$^{  }$ & 2 & 1.15 & -0.077 & 120.5  &  -40.4 &  25.9  &  -2.31 & -20.04 &  -5.47 & 12.5 &  4.2 &  2.7 &  6.9 &  3.0 &  1.7 \\
 79404$^{  }$ & 1 &-1.32 & -0.197 & 134.4  &  -28.3 &  41.6  &  -5.18 & -15.43 &  -6.86 & 15.2 &  3.2 &  4.7 &  6.9 &  2.4 &  2.3 \\
 81914$^{  }$ & 2 & 0.15 & -0.119 & 141.9  &  -44.0 &   8.1  &  -6.73 & -14.66 &  -3.91 & 18.0 &  5.6 &  1.0 &  1.8 &  2.0 &  0.6 \\
 82545$^{  }$ & 2 &-2.47 & -0.223 & 153.6  &  -37.7 &  10.7  &  -3.26 & -19.78 &  -3.70 & 20.9 &  5.1 &  1.5 &  0.9 &  2.7 &  0.8 \\
 84970$^{  }$ & 1 &-2.95 & -0.223 & 171.6  &    1.4 &  19.7  &  -1.37 & -20.10 &  -5.05 & 20.5 &  0.2 &  2.3 &  3.6 &  2.3 &  1.0 \\

\hline
\end{tabular}\\
$^1$multiple, $^2$emm.l./variable, $^3$variable, $^4$rotnl. variable,
$^5$Be, $^6$emm.l./binary, $^7$Be/neb.emm., \\
$^9$variable/$\beta$Cep, $^{10}$binary, $^{11}$ecl.binary,
$^{12}$variable/$\beta$Cep, $^{13}$spec.binary, $^{15}$double, \\
$^{16}$variable/$\beta$Cep/double?,
$^{18}$ellips.variable/double? \\
\end{minipage}
\end{table*}
\begin{table*}
\hspace*{-3.5cm}
 \centering
 \begin{minipage}{140mm}
  \caption{HIP-nos. of the remaining selected stars}
  \begin{tabular}{@{}rrrrrrrrrr@{}}
  \hline
   2484 &   2505 &   5566 &   7588 &   7943 &   8886 &  10602 &  10944 &  11249 &  12692 \\
  15338 &  15404 &  15444 &  15627 &  15770 &  15988 &  16147 &  16210 &  16244 &  16470 \\
  16611 &  16803 &  17499 &  17531 &  17563 &  17573 &  18033 &  18190 &  18213 &  18216 \\
  18788 &  19860 &  20042 &  20063 &  20171 &  20186 &  20884 &  21192 &  21281 &  22109 \\
  23607 &  23767 &  24244 &  25813 &  26248 &  26487 &  26623 &  26640 &  29426 &  30069 \\
  30122 &  30675 &  31278 &  31362 &  31685 &  32677 &  32912 &  33015 &  33579 &  34045 \\
  35054 &  35785 &  36188 &  37304 &  38455 &  38863 &  39138 &  39360 &  39906 &  40581 \\
  42177 &  42637 &  43105 &  43394 &  43878 &  43937 &  45080 &  45418 &  45941 &  46283 \\
  47119 &  47391 &  47452 &  51437 &  51576 &  52370 &  52419 &  52502 &  52701 &  52736 \\
  54767 &  55597 &  58484 &  60000 &  60718 &  60855 &  61789 &  62786 &  65474 &  66821 \\
  67301 &  68269 &  69389 &  71353 &  76243 &  76669 &  78493 &  79653 &  82673 &  82902 \\
  83895 &  85391 &  85792 &  86414 &  89482 &  89908 &  90200 &  90422 &  91235 &  91729 \\
  92614 &  92855 &  93104 &  93187 &  93231 &  93299 &  93805 &  95400 &  95951 &  96052 \\
  96417 &  96468 &  97376 &  98412 &  98754 & 100751 & 101017 & 101421 & 101475 & 101716 \\
 101746 & 101868 & 103089 & 103532 & 103616 & 104105 & 105148 & 105282 & 106604 & 107462 \\
 107664 & 107930 & 108022 & 109139 & 112781 & 115990 & 116805 &        &        &        \\

\hline
\end{tabular}
\end{minipage}
\end{table*}

\bsp

\label{lastpage}

\end{document}